\documentclass[10pt]{article}
\usepackage [cp1251] {inputenc}
\usepackage{amsfonts,amsmath,amsxtra}

\newcommand{\beq}{\begin{equation}}
\newcommand{\eeq}{\end{equation}}
\newcommand\bqa {\begin{eqnarray}}
\newcommand\eqa {\end{eqnarray}}
\newcommand{\bec}{\begin{cases}}
\newcommand{\eec}{\end{cases}}
\newcommand{\bei}{\begin{itemize}}
\newcommand{\eei}{\end{itemize}}
\newcommand{\bee}{\begin{enumerate}}
\newcommand{\eee}{\end{enumerate}}

\newcommand\pr {\partial}
\newcommand{\fr}{\frac}

\newcommand\nn {\nonumber}

\newcommand{\bear}{\begin{array}}
\newcommand{\enar}{\end{array}}

\newcommand{\D}{\mathcal{D}}

\newcommand{\bracket}[3]{\Bigl\langle #1 {\Bigr| #2  \Bigl|} #3 \Bigr\rangle}

\begin{document}

\def\I{{\rm i}}

\def\h{\hbar}

\def\t{\theta}
\def\T{\Theta}
\def\w{\omega}
\def\ov{\overline}
\def\a{\alpha}
\def\b{\beta}
\def\g{\gamma}
\def\s{\sigma}
\def\l{\lambda}
\def\wt{\widetilde}
\def\t{\tilde}

\vspace{-6.0cm}

\begin{center}
\hfill ITEP/TH-75/09\\
\end{center}

\vspace{1cm}

\centerline{\bf \Large A simple way to take into account back reaction on pair creation}

\vspace{10mm}

\centerline{\bf Emil T. Akhmedov\footnote{akhmedov@itep.ru} and Philipp Burda\footnote{blade131@yandex.ru}}

\vspace{5mm}

\centerline{ITEP, B. Cheremushkinskaya, 25, Moscow, Russia 117218}

\centerline{and}

\centerline{MIPT, Dolgoprudny, Russia}

\date{\today}

\begin{abstract}
We propose a simple and systematic way of accounting for the back reaction
on the background field due to the
pair creation in the four--dimensional scalar QED.
This method is straightforwardly generalizable to the gravity
backgrounds. In the case of QED with the instantly switched on constant electric field background we obtain
a remarkably simple formula for its decay rate.
\end{abstract}


\section{Introduction}

Back reaction on classical background fields due to the quantum pair creation is an important
problem for the understanding of the black hole and de Sitter space thermodynamics. The situation with the de Sitter space is the most controversial \cite{Polyakov:2007mm}--\cite{Volovik:2008ww}.
At any rate, the proper way of taking into account the back reaction should be a first
step towards quantization of gravity. Apparently the moment when the
back reaction becomes relevant for the gravity background
coincides with the moment when the gravity becomes quantum.

All the existing ways of taking into account back reaction are rather complex
and do not allow for a clear study of the situation. For that reason it is obviously tempting to
analyze the phenomenon in a simple setting of the four--dimensional
scalar QED with the electric field background,
where the situation is supposed to be clear at least qualitatively. Even in the latter case
the standard ways of accounting for the back reaction are rather involved (see e.g. \cite{TsamisWoodard} and \cite{Cooper:1989kf}). The goal of this note is to propose a simple
way to calculate the decay rate of the background electric field, which can be straightforwardly
generalized to the gravity backgrounds.

Let us formulate the problem here. We assume that somehow at the moment of time $t=0$
a constant (everywhere in space) electric field was created. Saying another way,
we neglect the boundary effects. We assume that at $t=0$ there
are no electrically charged particles present on top of the field. At this moment one turns on
interactions and the pair creation begins \cite{Schw} (see as well \cite{Gitman1} for the further
development and \cite {Affleck:1981bma}--\cite{Russo:2009ga} for the more recent progress and generalizations). On general physical grounds one expects that the value of the field will be decreasing
at least due to the work performed to create pairs and to accelerate them. Due to the symmetry
of the problem the decay should proceed homogeneously is space.

The standard way to take into account the back reaction, considering the background field as classical,
is to calculate the electric current which is created and to take into account the field due to this current \cite{TsamisWoodard}, \cite{Cooper:1989kf}. If the field is directed along $z$ axis one has to consider
the Heisenberg evolution of the $J_3$ component of the electric current up to some moment of time $t$ and average it over the initial state of the problem. This way one would find the value of the created current
at the moment $t$. Because the initial state is unstable one has to use the Keldysh--Schwinger diagram technic rather than the Feynman one \cite{TsamisWoodard}. Substituting the resulting value of the current into the RHS of the Maxwell's equation, one can find the field produced by it, which reduces the value of the original background field \cite{TsamisWoodard}.

Such a procedure is rather involved and becomes even harder when generalized to the gravity backgrounds.
In this note we propose another obvious and equivalent to the above mentioned one method
to solve the problem in question. We propose to consider the Heisenberg--Euler effective action (including the Schwinger's imaginary contribution) for the background electric field in the scalar QED  \cite{Cooper:1989kf}, which is switched on at $t=0$.
Obviously to obtain the decay rate of the background electric field one just has to solve the equations of motion following from the effective action. Along this line we find an unexpectedly simple answer (see (\ref{answ}) below). Our idea is to obtain the decay rate
at the initial stage of the process and to approve that it is small for the
subcritical values of the background electric field and as well to see whether
the rate is big for the overcritical values of the electric field. Our eventual goal
is of cause to find the decay rate for the overcritical field, because it is such a value
of the field which is interesting
from the perspective of the generalization of our considerations to the gravitational
backgrounds.

\section{Effective action and equations of motion at one loop}

We would like to solve the following equation:

\bqa\label{Maxwell}
\partial^\mu F_{\mu\nu}(x,t) = \left\langle in \left| \hat{j}_{\nu}(x,t)\right|in \right\rangle
\eqa
in the scalar QED. On the RHS of this equation stands the current due to the created pairs, which is
calculated as the average of the Heisenberg evolution
of the current operator $\hat{j}_\mu$ over the initial state $|in\rangle$ \cite{TsamisWoodard,Cooper:1989kf}. One can straightforwardly convert
this expression into the interaction picture, where in our case the interaction part of
the Hamiltonian is just the interaction of scalars with the classical electromagnetic field.
Due to spacial homogeneity of the problem in question this
average does not depend on $x$.

The initial state $|in\rangle$ in question is the exact free theory vacuum state
for the scalars in the constant electric field background. If such a state was
an eigen--state of the Hamiltonian, i.e. if $\langle out| \hat{S} |in\rangle$, with $\hat{S}$
being the S--matrix operator, was a pure phase, then we could have made the following transformations:

\bqa
\left\langle in \left| \hat{j}_{\nu}(t)\right|in \right\rangle = \left\langle in \left| \hat{S}^{-1} T\left[\hat{j}^{(0)}_{\nu}(t)\, \hat{S}\right]\right|in \right\rangle
= \left\langle in \left|\hat{S}^{-1}\right|out\right\rangle \, \left\langle out\left|T\left[\hat{j}^{(0)}_{\nu}(t)\,\hat{S}\right]\right|in \right\rangle,
\eqa
where $\hat{j}_{\nu}^{(0)}(t)$ is the current operator in the interaction picture.
Along these lines we would arrive at the standard functional integral in the in--out formalism.
However, in our case the in--state is not an eigen--state of the Hamiltonian for all times and, hence, $\langle out| \hat{S} |in\rangle$ is not
a pure phase. Hence, we have to apply a different procedure due to Schwinger \cite{Schw}.

According to \cite{Jordan:1986ug} we have to introduce the generating functional:

\bqa
e^{\I\,W[A^+,\,A^-]} \equiv \sum_\alpha \left\langle in \left|\hat{S}^{-1}\right|out,\alpha\right\rangle_{A^-}\,
\left\langle out, \alpha \left|\hat{S}\right|in\right\rangle_{A^+},
\eqa
which involves a summation over a complete set of out--states. In our case, $A_{\pm}$
are gauge fields and, at the same time, sources for the currents $\langle in |\hat{j}_\mu(t)|in \rangle = \delta W[A^-,\, A^+]/\delta A^+_\mu(t)|_{A^+=A^-=A} = - \delta W[A^-,\, A^+]/\delta A^-_\mu(t)|_{A^+=A^-=A}$. Note that the effective action
obeys the conditions \cite{Jordan:1986ug}: $W[A,\,A] = 0, \quad W[A^-,\,A^+] = - W[A^+,\,A^-]^*$.

The effective action in question for the background classical constant electric field in the four--dimensional scalar QED is equal to:

\bqa\label{Euclefac}
\I\,W[A^-,\,A^+] = \nonumber \\ \log \left[\int \D{\phi^*}(x) \D{\phi}(x) \; \int_{\phi_-(t=0,\vec{x})=0} \D{\phi^*_-}(x,t) \D{\phi_-}(x,t) \; \exp\left\{ \frac{\I}{2} \,\int_0^{+\infty} dt \int d^3\vec{x}\, \left(\left|D^{-}_{\mu}\phi_-\right|^2 + m^2 \, \left|\phi_-\right|^2 \right) \right\}\times\right. \nonumber \\ \times \left.\int_{\phi_+(t=0,\vec{x})=0} \D{\phi_+^*}(x,t) \D{\phi_+}(x,t) \; \exp\left\{ - \frac{\I}{2} \,\int_0^{+\infty} dt \int d^3\vec{x}\, \left(\left|D^{+}_{\mu}\phi_+\right|^2 + m^2 \, \left|\phi_+\right|^2 \right) \right\}\right].
\eqa
In this equation we have to impose $\phi_-(t=+\infty, x)=\phi_+(t=+\infty,x)=\phi(x)$ at future infinity and assume the single common integration over this value of the field at the future infinity \cite{Jordan:1986ug},
$\phi_{\pm}$ are the scalar fields for the direct and reverse time directions, $D^{\pm}_\mu = \partial_\mu - i\, e\, A_\mu^{\pm}$.

With the use of the effective action under consideration, one can rewrite the equation (\ref{Maxwell}) in the following way

\bqa\label{MainEq}
&& \pr^{\mu} F_{\mu\nu} = \I\, \int_{\phi(t=0,\vec{x})=0} \D{\phi^*} \D{\phi} \; \phi^*\! \left( \I e {\mathop{\pr}^{\leftrightarrow}}_{\nu} + 2 e^2 A_{\nu} \right)\! \phi\;  \exp\left\{ - \frac{\I}{2} \, \int_C dt \int d^3x \, \phi^*\!\left(-D_{\mu}^2 + m^2 \right) \phi \right\} = \nn\\
&& = \I \left.\left[ \I e \left(\fr{\pr}{\pr{y^{\nu}}} - \fr{\pr}{\pr{z^{\nu}}}\right) + 2 e^2 A_{\nu} \right] \int_{\phi(t=0,\vec{x})=0} \D{\phi^*} \D{\phi} \; \phi^*\!(y_\mu)\phi(z_\mu) \;  \exp\left\{ - \frac{\I}{2} \, \int_C dt \int d^3x \, \phi^*\!\left(-D_{\mu}^2 + m^2 \right) \phi \right\}\right|_{y=z=x}
\label{eom1}
\eqa
In this equation we have unified $\phi_{\pm}$ into one field $\phi$ and according to (\ref{Euclefac}) we have extended the time axis over the Keldish--Schwinger type contour $C$ consisting of the two parts --- $C_+$ going from $0$ to $+\infty$ and $C_-$ going backward.

In the last line of (\ref{eom1}), under the action of the differential operator, we have the in--in Keldish-Schwinger ``++'' type propagator on the half of $R^{3,1}$. We would like to express this in--in propagator for the half of $R^{3,1}$ ($t \in [0,+\infty)$) through the in--out propagator for the full $R^{3,1}$ ($t\in (-\infty,+\infty)$). First, to express the in--in propagator, $G_{in-in}(y,z)$, through the in--out one, $G_{in-out}(y,z)$, for the full $R^{3,1}$ ($t\in (-\infty,+\infty)$), recall their definitions: $G_{in-in}(y,z) = \langle in|T \phi(y)\,\phi(z)|in\rangle = \sum_\lambda f^*_{in,\lambda}(y)\,f_{in,\lambda}(z)$ and $G_{in-out}(y,z) = \langle out|T \phi(y)\,\phi(z)|in\rangle/\langle out|in\rangle = \sum_\lambda f^*_{out,\lambda}(y)\,f_{in,\lambda}(z)\,\langle out|b_{out,\lambda} \, a^+_{in,\lambda}|in\rangle/\langle out|in\rangle$. In these expressions we have expanded the field $\phi$ in terms of the in--harmonics $f_{in,\lambda}$ and out--harmonics $f_{out,\lambda}$, which are properly normalized and correspond to the eigen--values $\lambda$. Using Bogolyubov transformation from the out--
to the in--harmonics one can express $G_{in-in}(y,z) = G_{in-out}(y,z) + G^*_{in-out}(y,z) = 2 {\rm Re} \left[ G_{in-out}(y,z) \right]$. And finally, in order to express the in-out propagator for the full $R^{3,1}$ space through the in-out propagator for the half-space $t \in [0,+\infty)$ with Dirichlet boundary conditions one just need to use mirror sources.

To proceed we need to convert all the expressions into the Euclidian signature.
We are going to make the Wick rotation back to the Minkowski space--time ($x_0=\I\, t$) at the end
of the calculation. Using the considerations of the previous paragraph, it is straightforward to see that in Euclidian space the equation (\ref{MainEq}) can be rewritten as:

\bqa
\pr_{\mu} \, F_{\mu\nu} =
- 2 \,\left. \left[ \I e \left(\fr{\pr}{\pr{y^{\nu}}} - \fr{\pr}{\pr{z^{\nu}}}\right) + 2 e^2 A_{\nu} \right] {\rm Re} \left[\bracket{z}{\fr{1}{-D_{\mu}^2 + m^2}}{y} - \bracket{z}{\fr{1}{-D_{\mu}^2 + m^2}}{\bar{y}}\right]\right|_{y=z=x}.
\label{eom}
\eqa
The point $\bar{y}$ is the position of the mirror source symmetric to $y$ with respect to the surface $x_0=0$.

The RHS of this equation can be calculated if the explicit expression for the background field is given.
Hence, one has to look for the self consistent solution $A_3(x_0)$ of this equation. The solution
should be self consistent in the sense that the RHS of (\ref{eom}), being calculated for the
given $A_3(x_0)$, should lead to the same $A_3(x_0)$ as the solution of (\ref{eom}).
Unfortunately, this problem has been solved only numerically \cite{Cooper:1989kf}.

We would like to develop here the analytical approach in some approximation specified below. In particular, one can calculate exactly the RHS for the constant electric field background.
The picture is self consistent if, as the result of (\ref{eom}), the rate of the change of the background field is small. Otherwise one has to look for another approach. In any case our approximation works
in the classical limit, when $\hbar\to 0$ and the RHS of the equation in question is vanishingly small
due to its one loop origin.

So we assume that our background electric field is slowly changing in time and substitute into
the LHS of (\ref{eom}) $F_{\mu\nu} = - \I E(x_0) \delta_{3\left[\mu\right.}\delta_{\left.\nu\right]0}$ (imaginary unit is due to the Euclidean version of the theory). At the same time we neglect the time dependence of the field on the RHS, i.e. we put $A_{\mu}= E x_0 \delta_{\mu 3}$ into the propagator. Then the equation in question reduces to

\beq
\fr{d{E(x_0)}}{d{x_0}} = 2 \left. \left[\I e \left(\fr{\pr}{\pr{y_{3}}} - \fr{\pr}{\pr{z_{3}}}\right) + 2 \, e^2\, E\, x_0\right] {\rm Re}\left[\bracket{z}{\fr{1}{-D_{\mu}^2 + m^2}}{y} - \bracket{z}{\fr{1}{-D_{\mu}^2 + m^2}}{\bar{y}}\right]\right|_{y_\mu=z_\mu=x_\mu}.
\label{eomS}
\eeq
In the case under consideration the RHS can be computed exactly.
It is worth stressing at this point that the result of our calculation is gauge and Lorentz invariant because eq. (\ref{eom}) is written down in a gauge and Lorentz invariant way.

We proceed with the computation of the propagator of charged scalars in the given external
gauge potential:

\bqa
\bracket{z}{\fr{1}{-D_{\mu}^2 + m^2}}{y} & = & \bracket{z}{ \int_0^{\infty} d{T} e^{-\left(-D_{\mu}^2 + m^2\right)T} }{y} = \int_0^{\infty} d{T} e^{-m^2 T} \bracket{z}{e^{D_{\mu}^2 \, T}}{y} = \nn\\
& = & \int_0^{\infty} d{T} e^{-m^2 T} \int_{x(0)=y; x(T)=z} \D{x(\tau)} \; e^{-\int_0^T (\fr{1}{4}\dot{x}^2 +\I e A_{\mu}\dot{x}_{\mu}) d\tau }.
\label{propertimePI}
\eqa
Then, in our case
\beq
\bracket{z}{\fr{1}{-D_{\mu}^2 + m^2}}{y} = e^{-S_{cl}[z,y]} \int_0^{\infty} d{T} e^{-m^2 T} \fr{eET}{(4\pi T)^2 \sinh(eET)},
\eeq
where

\beq
S_{cl}[z,y] = \fr{(y_1-z_1)^2 + (y_2-z_2)^2}{4T} + \fr{e\,E\, \coth(e\,E\,T)}{4}\left[(y_0-z_0)^2 + (y_3-z_3)^2\right] + \I\fr{eE}{2}\,(z_0 + y_0)\,(z_3 - y_3).
\eeq
is the classical action of the charged particle in the gauge field background in question.
In the case of the propagator with $y\to \bar{y}$ one has to change only

\beq
S_{cl}[z,\bar{y}] = \fr{(y_1-z_1)^2 + (y_2-z_2)^2}{4T} + \fr{e\,E\, \coth(e\,E\,T)}{4}\left[(y_0+z_0)^2 + (y_3-z_3)^2\right] + \I\fr{eE}{2}\,(z_0 - y_0)\,(z_3 - y_3).
\eeq
Returning to the Minkowski space--time via the replacements $x_0 \rightarrow \I t$ and $E \rightarrow \I E$, we get

\beq
\fr{d{E}}{d{t}} = - {\rm Re}\left(\I \fr{e^3 E^2 t}{8\pi^2}  \int_0^{\infty} d{T} e^{-m^2 T} \fr{1}{T\sin(eET)}\right).
\label{eomLG}
\eeq
On general physical grounds one can expect only such a contribution to the RHS of (\ref{eomLG}), which originates from the imaginary part of the Heisenberg-Euler Lagrangian describing the pair creation.

Now, using that

\beq
{\rm Im} \left( \int_0^{\infty} d{T} e^{-m^2 T} \fr{1}{T\sin(eET)} \right) = - \ln \left(1+e^{-\fr{m^2\pi}{eE}}\right),
\eeq
we obtain

\beq
\boxed{
\fr{d{E}}{d{t}} = - \fr{e^3 E^2 t}{4\pi^2}  \ln \left(1+e^{-\fr{m^2\pi}{eE}}\right)
}
\label{answ}
\eeq
This formula gives us one loop exact answer for the decay rate of the background electric field in our approximation. Ironically we find the exact decay rate of the field in the approximation when the field is held constant, which of cause is just a partial solution of the problem in any case. The generalization of our formulas to any dimension is straightforward.

There are several points which are worth mentioning here. First,
if we were considering the effective action in question in the full $R^{3,1}$,
as it was done in the original Schwinger's setting,  rather than
in the $t>0$ half of $R^{3,1}$, the RHS of (\ref{Maxwell}) would vanish,
because of its time reversal ($t\to -t$) invariance.
Second, it is straightforward to calculate that the VEV $\langle J_0 \rangle$ in our setting
does vanish as it should be, because the charges are created in pairs.
Third, it is as well easy to observe that if one were considering
the fate of the constant magnetic field instead of the electric one, he would
have found that the field would remain constant, because the corresponding current does vanish.
Which should be the case
due to the absence of the imaginary contribution to the effective action in magnetic field background.

\section{Discussion and conclusions}

The decay rate (\ref{answ}) looks suspicious because of the dependence of its RHS on $e^3$.
One might think that as we change from particles to antiparticles the RHS changes the sign and, as the result, the background field grows rather than decays. We should stress that the RHS
is always negative and depends on the modulus of the charge $e$.

To clarify the situation we would like to reobtain the answer (\ref{answ}) for the decay rate on the general physical grounds. Note that in our problem the initial density of charges in space is zero. Otherwise the space would have had non--zero conductance, $\sigma$, and the decay rate would have been exponential: $dE_z/dt = j_z = - \sigma \, E_z$.

Taking into account the set up of our problem, which is formulated at the end of the introduction
section, we can see that the decay of the field proceeds homogeneously.
Hence, the energy of the electric field per unit volume $E^2/8\,\pi$ is spent on the work performed by the
field on the creation of pairs. This work is proportional to $e\,E \,z = e\,E \,t$, taking into account that we set the light speed $c=1$. Here $z$ is the separation distance between the members of the pair reached during the observation time $t$.

Let us clarify this relation. In the gauge when $A_0$ is not zero, we have
$p_0 = \sqrt{m^2 + p_x^2 + p_y^2 + p_z^2} - e \, E\, z.$
Then, the distance between the turning points is
$\Delta z = \frac{2\,\sqrt{m^2 + p_x^2 + p_y^2}}{e\,E}.$ Similarly in the gauge when $A_z$ is not zero,
we have $p_0 = \sqrt{m^2 + p_x^2 + p_y^2 + (p_z-e\,E\,t)^2}.$
Hence, the separation time between the turning points is
$\Delta t = \frac{2\,\sqrt{m^2 + p_x^2 + p_y^2}}{e\,E} = \Delta z.$
This observation establishes the relation between the separation distance and the observation time.

Thus,

\beq
\frac{d}{dt} E^2 \propto - 2 e\, E\, t \, w(E),\label{TsWod}
\eeq
where $w(E)\propto e^2 \, E^2 \,e^{-\fr{m^2\pi}{eE}}$ is the approximate Schwinger's pair creation probability rate per unit time and unit volume. Thus, we restore the leading approximation of our exact one loop result (\ref{answ}). It is worth stressing at this point that somewhat similar formula (in the light cone frame) to (\ref{TsWod}) was obtained in \cite{TsamisWoodard}.

The result (\ref{answ}) confirms the obvious expectation, that weak electric field ($eE << m^2$) changes slowly in time, because in this limit $ \ln\left(1+e^{-\fr{m^2\pi}{eE}}\right) \rightarrow 0$ and therefore $d{E(t)}/d{t} \rightarrow 0$. Note as well that this result is applicable in case only when the
decay rate is small, which is valid for the short period of time
$t << (m/e\, E)\, e^{\frac{\pi \, m^2}{e\,E}}$.
On the other hand, it gives us a hint about strong field limit, because naive (and illegal) application of this formula in the strong field limit ($eE >> m^2$) tells us that there will be a fast decay $E(t) \propto 1/t^2$. I.e. in the case of the overcritical field we can not apply our approximation.

How then one can find the back reaction rate in the case of the overcritical field?
Should one still consider the background field as classical? Will the decay rate still be
non--analytic in the value of the background field, if the field is grater than the critical value?
The reason why we are interested in the overcritical electric field background is that it is the
simpler model example predecessing the consideration of the gravity backgrounds, where the overcritical
background field is the most interesting case.

\section{Acknowledgments}

We would like to acknowledge valuable discussions with D.Diakonov, D.Galakhov, S.Gavrilov, A.Gorsky, D.Krotov, V.Losyakov, A.Monin, A.Morozov, I.Polyubin, A.Rosly, A.Roura, C.Schubert, V.Shevchenko and A.Zayakin. Our work is partly supported by RFBR grant 08-02-00287 (Ph.B.), by the Russian President's Grant of Support for the Scientific Schools NSh-3036.2008.2(Ph.B.), by the Dynasty Foundation (Ph.B.) and by Federal Agency for Science and Innovations of Russian Federation (contract 02.740.11.5029) (Ph.B.).

\end{document}